# EXPANDING VERSUS NON-EXPANDING UNIVERSE

Antonio Alfonso-Faus
Escuela de Ingeniería Aeroespacial
E.U.I.T. Aeronáutica, Universidad Politécnica de Madrid
e-mail: aalfonsofaus@yahoo.es
Plaza Cardenal Cisneros, 3
28040 Madrid, Spain

**Abstract**

In cosmology the number of scientists using the framework of an expanding universe is very high. This model, the big-bang, is now overwhelmingly present in almost all aspects of society. It is the main stream cosmology of today. A small number of scientists are researching on the possibility of a non-expanding universe. The existence of these two groups, one very large and the other very small, is a good proof of the use of the scientific method: it does not drive to an absolute certainty. All models have to be permanently validated, falsified. Ockham's razor, a powerful philosophical tool, will probably change the amount of scientists working in each of these groups. We present here a model where a big-bang is unnecessary. It ends, in a finite time, in a second INFLATION, or a disaggregation to infinity. We also discuss the possibilities of a non-expanding universe model. Only a few references will be cited, mainly concerned with our own work in the past, thus purposely avoiding citing the many thousands of professionals working in this field.

# 1 Introduction

The scientific approach to cosmology (Galileo´s scientific methodology) started with the general relativity theory of Einstein. He applied his field equations to the case of a homogeneous and isotropic universe (the cosmological principle), arriving at his two well known cosmological equations. From Newton gravitation was known to be an attractive force, due to the presence of mass, and a kind of "collapsing" force was a natural explanation for local physics (falling of bodies on earth, mass concentration in celestial systems, planetary and satellite orbits in dynamical equilibrium and so on). Nevertheless there was no evidence of an overall collapsing universe. Looking at the sky a static, non expanding universe, was a natural and intuitive picture. After all, the moon was not falling on the earth, and a sort of dynamical equilibrium was the accepted view for that. Therefore the whole universe could be in that state too.

The theoretical problem of a gravitationally collapsing universe, something that most probably should have occurred in the past, was also present in the initial Einstein´s cosmological equations. The problem was solved by Einstein introducing everywhere an outward pressure, a pushing force equilibrating the gravitational pull. Mathematically this was done by Einstein adding a cosmological constant $\Lambda$ to his equations. Then one could have a static universe as was the general belief at that time.

This solution to the theoretical problem of a universal collapse was in trouble very soon. The Hubble red-shift observations were known during the 1920´s: light from distant galaxies was observed to be red shifted more and more as one could observe them deeper and deeper in space. Now, interpreting these observations as a kind of "Doppler effect", the galaxies were seen as receding from us faster and faster going deeper in space. The conclusion was that the universe is expanding, a conclusion that immediately had rapid and wide acceptance.

The rising doubts on the stability of the Einstein static universe gave more support to the expansion idea. If the universe was in an unstable state, there could be expansion or contraction. The Hubble observation clearly pointed towards the expansion alternative.

We know that the laws of physics are reversible in time. If the universe is expanding with cosmological time this strongly supports that, on the average, the density and temperature of the whole universe are decreasing: going back in cosmological time this means a higher temperature and density in the past. But, where are we going to stop? Or, reversing the time again, how are we going to begin?

In mathematics one deals with infinities and its inverses, the zeros, without any problem. They are usually encountered in limiting processes. Nature has other kind of reality. Many times the infinities are avoided by some physical process, like viscosity. For example, when we have discontinuities in the velocities, like in an aero dynamical foil in the trailing edge of the aircraft´s wing, or when a stream of some fluid has to turn around a very sharp corner: viscosity enters into the picture and the problem is solved. The same happens with the zeros, for example the case of the vacuum concepts. It is evident that an infinite density and temperature at the "beginning" of the universe (~ zero time) seems unnatural and unphysical. Quantum mechanics has to enter into the picture.

General relativity deals with well behaved continuous variables. Nevertheless mathematical "singularities" do occur. Then the physical treatment may be to use quantum mechanics, to introduce discreteness. By doing so we can avoid the initial singularities in the universe. We may think of an initial quantum of time (just a quantum of space divided by the speed of light c) or a quantum of energy (Planck´s constant divided by the quantum of time). The application of the Heisenberg uncertainty principle solves the problem of singularities. From a physical point of view the duality of nature, continuous (wave-like) and discrete (particle-like) has to be tied to a continuous growth of the number of quanta (e.g. replication) keeping at the same time in some way a universal range of interaction (any part of the universe connected with the rest of the universe).

To think of the birth of the universe out of an infinite density and temperature, a big-bang as Hoyle unwillingly coined it, is really unnecessary. This is due to the application of quantum mechanics. Unfortunately we still do not have a workable quantum theory of gravity, and probably we will never do unless we change the track. We think that to take general relativity, in a sense, as an "absolute truth" and at the same time take quantum mechanics as another "absolute truth" in itself, makes their junction impossible.

Something of this sort is happening when dealing with the cosmological constant $\Lambda$ from two different points of view: approaching the evaluation of this parameter from a cosmological point of view, the reason of its birth, gives a factor of about $10^{122}$ difference as compared with the approach from the standard model of fundamental particles. It seems clear that we should not mix scales. The scaling factor between Planck´s world and the universe is about $10^{61}$. We have found that Planck´s constant [1] and the $\Lambda$ constant vary in an inverse way as the square of the scale. If one keeps the physical laws within each scale, without mixing them, each scale has a different Planck´s constant and a different $\Lambda$ constant, their product being always of order one. We have to consider each scale as a universe in itself and do not mix them. Then the contradiction just disappears [2].

**2 Expanding versus non-expanding universe.**

The central point here is: The scale of the present universe is the result of the expansion of an initial one (with different scale), like expanding a quantum black hole (Planck´s type), or is it in a kind of steady state?

The steady state model of the universe proposed more than 40 years ago by Hoyle, Bondi and Gold was abandoned because it had no evolution. This is a very important property that we observe in our universe, well known since decades. A static solution for the universe, with evolution, could be an alternative to the expanding one (big-bang). Understanding a static universe as a non expanding one means that the scale of the universe must be constant. Then the cosmological scale factor $a(t)$ must be a constant too, as well as all its derivatives that obviously should be zero. The question now is how to interpret the Hubble´s red shift observations. One way is to analyze the possibility that the speed of light changes with time [3]. This would certainly imply evolution. Looking at the dimensions of speed, a length divided by time, a simple assumption is to say that lengths L are constant, like the cosmological scale factor, and that time t grows so that the speed of light c ~ L/t varies inversely proportional to time. Of course the same must be true for any speed: this is to be expected in order to preserve the results of the special theory of relativity. They imply that for any speed $v$ the ratio $v/c$ must be constant.

A non expanding universe must imply that any momentum of any mass $mv$ must be constant. The conservation of momentum, in the absence of perturbations, is a well proven fact. On the other hand, general relativity predicts that momentum must be inversely proportional to the cosmological scale factor [9] $a(t)$

$$mv \sim 1/a(t) \qquad (1)$$

This is also a direct result of the "cosmic box" argument presented by Harrison [9]. Then a non-expanding universe, $a(t) = constant,$ implies the constancy of momentum, which is a Newtonian law, and a plausible consequence of the homogeneity of space too.

A flat universe, as observed in many instances, if it is non-expanding gives us a picture as Newton philosophically pointed out: it must be infinite in size. If space and time are absolute, in the Newtonian sense, and the speed of light is constant and finite, following the Michelson-Morlay experimental results, then as time runs there is more and more observable universe, ct. But no new galaxies are observed to enter in our visible universe, nowhere. One way to solve this problem, and to explain Hubble´s observation at the same time, is to consider the speed of light varying inversely proportional to time t [3]. Then the product *ct* is constant and we observe always the same universe. This picture solves too the horizon problem in cosmology, and close to the initial stages the speed of light should be huge. The condition *a(t) = constant* together with *v ~ 1/t* gives from (1) *m ~ t*. This is the mass-boom effect presented elsewhere [3]. The resultant picture seems coherent: non expanding universe, *a(t) = constant)*, Hubble red shift law fulfilled with the speed of light c ~ 1/t, evolution ensured, flat universe, mass-boom *m ~ t* which implies creation pressure [10], [11], and no need of the lambda constant, etc. There is another interesting argument: in order to keep the validity of the derivation of the Einstein´s field equations, out of an action principle [4], one needs the constancy of the ratio $G/c^3$. Then, a time varying c means a time varying G too. The point is that if one takes into account Mach´s principle in the form of equating any rest energy $mc^2$ to its gravitational potential energy, with respect to the mass M of the rest of the universe, one has

$$GM/c^2 \approx a(t) \approx ct \qquad (2)$$

which also is the condition for the universe to be a black hole. Taking the constant $G/c^3 = 1$ one has from (2)

$$M \approx t \qquad (3)$$

i.e. the mass-boom again [3]. It is interesting to note that in natural units G = c = 1 the result is exactly the same, $M \approx t$.

But, a big BUT, the speed of light is not observed to vary with time. The possibility of a very small and insignificant change with time, following an observation of a very small change of the fine structure constant [13], [14] is also questionable. The suggested time variation of c has no cosmological significance. Now, if we are forced to believe that the universe is flat, that the speed of light is constant and that the visible universe is always the same, then the universe *must be expanding*. The plausible condition may be that *a (t) ≈ ct = t* (and t = M). It looks like the best bet, as of today, is to think of a flat Euclidean universe, with constant speed of light c (that has been also taken as a constant by definition) and expanding with a cosmological scale factor close to *a(t) ≈ ct*, almost linearly with time at present.

**3 The cosmic box argument of Harrison [9]**

With the best bet we made above, the mass m of some fundamental particle has to be inversely proportional to *a(t)*: from the relation (1), with constant c and therefore constant v, one has the result

$$m \sim 1/a(t) \qquad (4)$$

This is a very interesting (and surprising) result. We do have a relation between masses of fundamental particles (in particular pions) and the age of the universe t through the Hubble "constant" H ≈ 1/t given by the Weinberg relation [16]

$$\hbar^2 H / (Gc) \approx m^3 \qquad (5)$$

In natural units, $G = c = \hbar = 1$ we get

$$m \approx 1/ t^{1/3} \qquad (6)$$

In natural units t is of the order of ~ $10^{61}$ so that m is of the order of

$$\sim 1/3 \ (10^{-20}) \ \text{Planck's mass} \approx 10^{-25} \text{ grams} \qquad (7)$$

i.e. a fundamental particle. But there is a more "fundamental" particle: the quantum of gravity [5].

**4 The quantum of gravity**

Since the gravitational field is not localized, so must be its possible quantum. Then if *a(t)* represents the size of the universe, and $m_g$ the equivalent mass of the quantum of gravity having momentum $m_g c$, the Compton wavelength of this quantum must be of the order of *a(t)*, i.e.

$$a(t) \approx \hbar/m_g c \qquad (8)$$

But this is precisely Harrison formulation (1). In natural units $c = \hbar = 1$ we get from (8)

$$m_g \approx 1/a(t) \qquad (9)$$

This is the smallest possible quantum of mass, and completely fulfills the "cosmic box" argument of Harris. And it does not conserve its momentum: it decreases with the size of the universe. The previous arguments of conservation of momentum [6] to imply a constant size universe (a non-expanding universe) are here seen to be invalid.

**5 Expansion with no big-bang**

We have to point out that the best bet; an expanding universe as shown in section 3, does not necessarily imply a big-bang as the origin of space-time. And much worse would be to try to imagine a big-bang occurring at "every point" of an infinite "vacuum": this would be an infinite "coincidence" very difficult to imagine, even to believe. There is no need for a big-bang: inflation suffices to do the job, and "inflates" (by a factor of about $10^{60}$) a Planck fluctuation (a quantum black hole) to a size close to the present size of the universe. And this is achieved in an extremely small time: a few tens of Planck´s time. The beginning of the universe, instead of an explosion as suggested by the big-bang name, may well be just a Planck´s fluctuation that inflated very rapidly indeed.

Some cosmologist, having a look at this review, may be thinking about the evidence from the cosmic microwave background radiation (CMBR), an indirect evidence for an initial big-bang, the CMBR being the relics of such "explosion". It might be so, time will probably tell. But it also may be that the CMBR is the relic of something else. In the next section we present an alternative to the big-bang origin of the CMBR.

**6 Millimeter black hole as the source of CMBR**

The origin of CMBR and dark matter can be explained with only one hypothesis [7]. A homogeneous distribution of millimeter black holes of 0.00635 cm could be the source of CMBR, as emission of such black holes. This emission is obviously blackbody as the observed CMBR. The same millimeter black holes will give the mass quantity which is usually stated as dark matter. Therefore, the effects explicated with dark matter could be stated with such black holes. The unknown nature of dark matter will then disappear. The distribution of such black holes would be the same postulated for dark matter.

**7 Strong evidence for an initial inflation**

We make an initial guess, an ANSANTZ approach, using the approximation $a(t) \approx ct$ which immediately gives for the speed of expansion of the universe $a´(t) \approx 1$. Now, the definition of the deceleration parameter $q$ is

$$q = - a´´(t)\, a(t) / (a´(t))^2 \approx - a´´(t)\, a(t) \approx - a´´(t)\, t \qquad (10)$$

There are three points which are very well approximated by many observations: close to the origin (very high red shift) the value of $q$ is about 0.5. At the present time the value of $q$ is about – 0.5. And we do know now that there is deceleration followed by acceleration with the turning point ($q = 0$) at about half the present age of the universe. We can fit a linear relation between $q$ and the dimensionless parameter $x = t/t_0$ where $t_0$ is the present age of the universe. Then the linear relation for $q$ is

$$q = ½ - x \qquad (11)$$

The differential equation from (10) and (11) is now

$$a''(t) = 1 - 1/2x \qquad (12)$$

We consider now the cosmological scale factor *a(t)* as dimensionless. This we do by referring it to the present cosmological scale value, taken as the unit reference. Then integrating (12) once we get

$$a'(t) = x - \tfrac{1}{2} \ln x \qquad (13)$$

where we see that for x = 1 we have *a'(t) = 1*, the choice stated at the beginning of this section. Integrating again (13) we get

$$a(t) = \tfrac{1}{2} x^2 - \tfrac{1}{2} (x \ln x - 1) \qquad (14)$$

where we see that for x = 1 we have *a(t) = 1*, the choice stated above. We plot now in the following figure 1 the cosmological scale factor from (14)

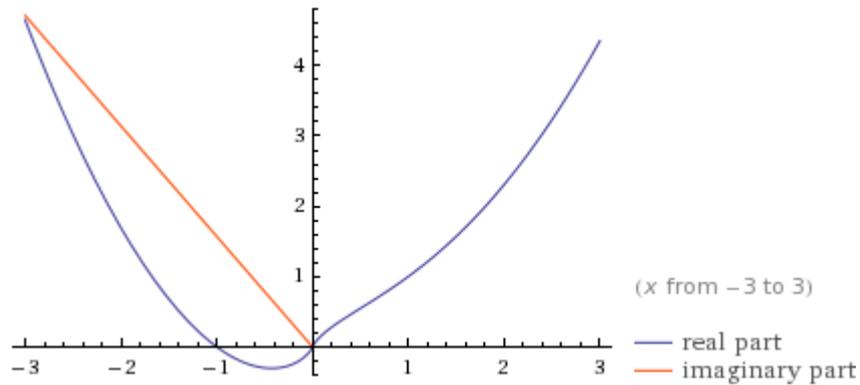

Fig. 1 Cosmological Scale Factor *a(t)* in terms of relative time x. The initial evident deceleration (x ≈ 0) is followed by acceleration after x = ½

## 8 Strong evidence of inflation followed by deceleration-acceleration. No evidence of a big-bang.

We plot now the speed of expansion given in (13) in the following figure 2 in terms of age x:

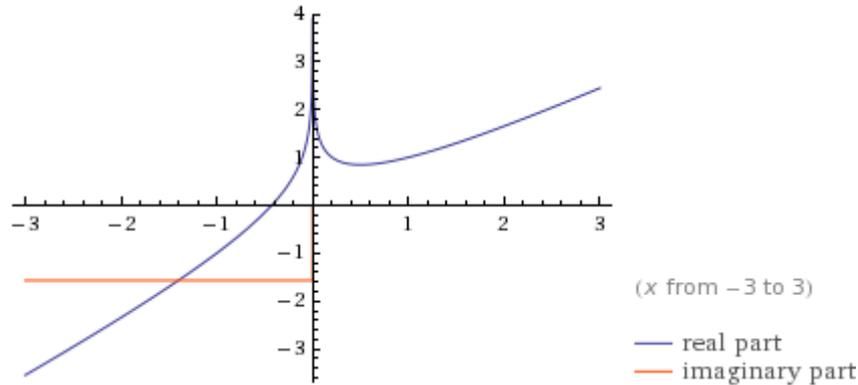

Fig. 2 Speed of expansion *a´(t)* in terms of relative time x. We see the initial infinite speed (x = 0) that is indicative of the initial inflation. Then deceleration is followed by acceleration

In this figure 2 there is no doubt that an infinite speed at the very beginning of the universe is clearly seen, and that this is also a very strong indication of an initial inflation.

The case for inflation is very well mathematically analyzed by using the definition of the Hubble parameter $H = a´(t)/a(t)$ (that has dimension 1/time) and the definition of *q* in (10) giving the differential equation,

$$H´ + (1 + q) H^2 = 0 \qquad (15)$$

An exponential expansion, the mathematical statement for inflation, may be expressed by the following relation

$$a(t) = \exp(Ht) \qquad (16)$$

that is in fact the definition of H. One gets the expression for H from the exponential expansion in (16) as

$$H = a´(t)/a(t) = (Ht)´ = H´t + H \qquad (18)$$

And from here we get

$$H' = 0 \quad i.e. \quad H = constant \qquad (19)$$

Then from (15) the numerical value of *q* that satisfies this condition is

$$q = -1 \qquad (20)$$

Hence, any value of x, the time or age of the universe, that allows the value of $q = -1$ is a strong indication of inflation. Looking at the figure 3 we see that inside the interval $(-\varepsilon, +\varepsilon)$, with $\varepsilon \ll 1$, the value of $H(x)$ runs from $-\infty$ to $+\infty$. Then it certainly goes through the inflation value. This is a very strong theoretical evidence for inflation: if we plot the Hubble parameter $H = a'(t)/a(t)$ (that has dimension 1/time) in terms of age x we get the following figure 3:

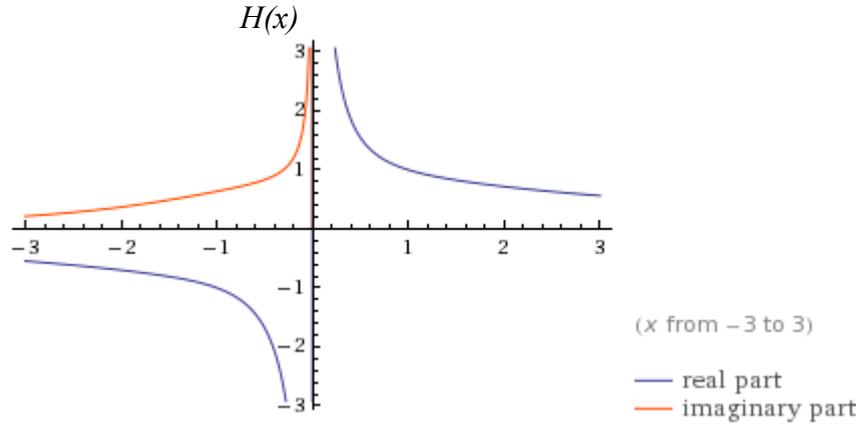

Fig.3. Hubble parameter *a´(t)/a(t)*: EVIDENT INITIAL INFLATION with infinite speed of expansion.

The plot of *H* in Fig. 3 is a conclusive theoretical evidence for an initial expansion as proposed by Guth [8] and Linde [12]. With this conclusion and the plot of the cosmological scale factor in Fig. 1 the theoretical non-expanding case for the universe is not supported in our present approach. This has been done without using the Einstein´s cosmological equations. Only the measured values for the acceleration parameter *q* have been considered here.

It looks like the initial inflation is enough to explain the birth and evolution of the universe afterwards. The well known name of a big-bang for an expanding model of the universe seems unnecessary. There is no need of a big bang as the beginning or birth of the universe. The inflation of an initial fluctuation, for example a quantum black hole of the Planck´s type, suffices to do the job.

## 9 A scientific modeling procedure to find the expansion factor

We have used three values of the deceleration parameter $q$ that are well known today. For very high red shifts the value of 0.5 seems to be a good approximation. There is some different values for the age at which the deceleration parameter is zero: they run from about 0.4 to about 0.7. For convenience of calculation we have taken the value x = 0.5 at which $q = 0$. It seems well established that before this point deceleration is present and $q$ is therefore positive. After this point we have acceleration (negative $q$) arriving at the present time with a value close to $q = - 0.5$. This are results from about ten years of observation from satellites. It is expected that in another 10 years we will know if the acceleration goes on, and at which rate. It may be that the value for another inflation, $q = - 1$, can be extrapolated in the future and then a disaggregating effect may be the end of the universe. The values involved for this possibility to occur are at about another $\sim 10^{10}$ years from now.

## 10 The physics behind the cause for expansion

### 10.1 Dark energy

The current figures for the baryonic matter content in the universe are close to 4%. For the non-baryonic dark matter the figure is about 6 times higher, about 24%. The remaining component, about 72%, is named as dark energy. This component, which is the most important one, is supposed to cause a positive pressure outwards. In this way this pressure counterbalance the gravitational attraction of the mass in the universe, and also pushes all other components outwards. This last effect is supposed to be responsible for the acceleration observed about ten years ago. The dark energy component thus solves the acceleration problem, as well as the presence of the 72% energy assumed, with some evidence, to be present in the universe. Needless to say that this dark energy has not been observed directly, and that its composition is totally unknown as of today. It is on the same foot as the dark matter issue.

### 10.2 The quantum vacuum pressure

A very old and worrying discrepancy is concerned with the value of the vacuum pressure, as compared with the cosmological outward pressure required to accelerate the universe. The discrepancy is of about 121 orders of magnitude. This is related to the standard model for fundamental particles. In cosmology the value of the outward pressure is proportional to the cosmological constant $\Lambda$, which is of the order of $10^{-56}$ cm$^{-2}$. This is the order of magnitude obtained by Zel´dovich in 1967 [16]. He derived an expression for the cosmological constant, the lambda constant $\Lambda$, based on quantum vacuum arguments, arriving at the approximate value

$$\Lambda \sim G^2 m^6/\hbar^4 \qquad (21)$$

where m is the mass of a fundamental particle, about $10^{-25}$ grams. If we use for m the Planck´s mass, about $10^{20}$ times greater, which means a factor of $\sim 10^{121}$, then we get the value obtained from the standard model for fundamental particles. The discrepancy can also be explained as a question of scaling Alfonso-Faus [2].

## 11 Conclusions

Today there is enough evidence to consider that the universe is flat, an Euclidean universe, with constant speed of light c. It is expanding with a cosmological scale factor close to *a(t) ≈ ct*, almost linearly with time. We have found unnecessary the widespread conviction that there was a big-bang at the very beginning of the universe. Quantum mechanics gives a very different view: a very small first quantum of everything (probably the quantum black hole represented by the Planck´s units). We have also found strong evidence for an initial inflation, followed by deceleration-acceleration. It may be that at about twice the present age of the universe a disaggregation to infinity could be possible. Finally the change of scale from the present universe to the Planck´s world is enough to explain the Λ $10^{121}$ discrepancy problems.

## 12 Acknowledgements

I am grateful to the owners of the Wolfram Mathematica Online Integrator that I have used to obtain the Figures 1, 2 and 3 for this work.